\documentclass[10pt,a4paper,english]{achemso}
\usepackage[T1]{fontenc}
\usepackage[latin9]{inputenc}
\usepackage{amsmath}
\usepackage{amssymb}
\usepackage{graphicx}
\usepackage{esint}

\makeatletter

\title{Geometric phase of light-induced conical intersections: Adiabatic
time-dependent approach }
\author{Gábor J. Halász(1), Péter Badankó(2) and Ágnes Vibók(2),(3)}
\affiliation{(1)Department of Information Technology, University of Debrecen,
H-4002 Debrecen, PO Box 400, Hungary }
\affiliation{(2)Department of Theoretical Physics, University of Debrecen, H-4002
Debrecen, PO Box 400, Hungary}
\affiliation{(3) ELI-ALPS, ELI-HU Non-Profit Ltd, H-6720 Szeged, Dugonics tér
13, Hungary}
\email{vibok@phys.unideb.hu}
\pdfpageheight\paperheight
\pdfpagewidth\paperwidth

\providecommand{\tabularnewline}{\\}

\usepackage{babel}

\usepackage{babel}

\makeatother

\usepackage{babel}
\begin{document}
\begin{abstract}
Conical intersections are degeneracies between electronic states and
are very common in nature. It has been found that they can also be
created both by standing or by running laser waves. The latter are
called light-induced conical intersections. It is well known that
conical intersections are the sources for numerous topological effects
which are manifested e.g. in the appearance of the geometric or Berry
phase. In one of our former works by incorporating the diabatic-to-adiabatic
transformation angle with the line-integral technique we have calculated
the Berry-phase of the light-induced conical intersections. 

Here we demonstrate that by using the time dependent adiabatic approach
suggested by Berry the geometric phase of the light-induced conical
intersections can also be obtained and the results are very similar
to those of the time-independent calculations.\end{abstract}
\begin{description}
\item [{Keywords:}] Born-Oppenheimer approximation; Conical intersections;
Light-induced conical intersections; Geometric phase; 
\end{description}

\section{Introduction}

Conical intersections (CIs) are degeneracies between two or more electronic
states and play an important mechanistic role in the nonadiabatic
dynamics of polyatomic molecules \cite{Born,Wigner,Aharonov1,Domcke1,Baer1,Graham1,Domcke2,Baer2}.
At the close vicinity of the CIs the Born-Oppenheimer adiabatic approximation
\cite{Born} breaks down due to the strong nonadiabatic coupling between
the nuclear and electronic motions. Several important photophysical
and photochemical processes like dissociation, proton transfer, isomerization
or radiationless deactivation of the excited states are associated
with the appearance of CIs \cite{Domcke2}. These degeneracies are
not isolated, rather they are connected points forming a seam and
can exist already between low lying states of triatomic molecule.
In truly large molecular systems they are very abundant. 

It is well-know that under ``natural'' (field-free) conditions CIs
cannot be formed between different electronic states in diatomic molecules
\cite{Wigner}. The one degree of freedom presents in these object
is generally not enough to span a branching space and therefore only
an avoided crossing result. However, applying standing \cite{Lenz1}
or running laser waves \cite{Lenz2}, CIs can be created even in diatomics.
In the first situation, the laser light induces CIs (``light-induced''
conical intersections, LICIs) which couple the center of mass motion
with the internal rovibrational degrees of freedom \cite{Lenz1}.
In the latter case, the rotational motion provides the missing degree
of freedom allowing the formation of a LICI \cite{Lenz2}. 

Recently, several theoretical and experimental studies have demonstrated
that similarly to the natural CIs the light-induced conical intersections
have also significant impact on the different dynamical properties
of molecules \cite{LICI1,LICI2,LICI3,LICI4,LICI5,LICI6,LICI7,LICI8,LICI9,LICI10}.
Among others it can strongly modify e. g. the spectra, the alignment,
the dissociation probability or fragment angular distribution of molecules
\cite{LICI1,LICI2,LICI3,LICI4,LICI5,LICI6,LICI7,LICI8,LICI9,LICI10}.
However, there are features in which the natural and light-induced
CIs differ significantly. As long as the position of a natural CI
and the strength of its nonadiabatic effects are inherent properties
of the electronic states of a molecule and are difficult to modify,
the energetic and spatial positions of the LICI can be controlled
by changing the parameter settings (intensity, frequency) of the laser
light. This latter can open up a new direction in the field of molecular
quantum control processes. 

The nonadiabatic couplings can become extremely large at the close
vicinity of the CIs. Numerous statical and dynamical nonadiabatic
phenomena originate from the presence of a CI. Longuet-Higgins and
Herzberg have demonstrated \cite{Longuet,Higgins}that each real adiabatic
electronic state changes sign when transported continuously along
a closed path encircling the point of CI. Mead and Truhlar conjuncted
this geometrical phase effect with the single electronic state problem
\cite{Mead1} and Berry generalized the theory \cite{Berry}, hence
the name Berry phase. Making sure that the electronic wave function
remains single valued one has to multiply it by a phase factor and,
as a consequence of it, this new electronic eigenfunction, instead
of being real, becomes complex. The fact that the electronic eigenfunctions
are altered has a direct impact on the nuclear dynamics even if it
happens in a single potential energy surface. Therefore having a nontrivial
Berry phase in a molecular system can be seen as a direct fingerprint
of the CI. 

In the last three decades several works made remarkable contributions
to the subject of molecular topological features \cite{Aharonov,Mukunda,englman1,englman2,englman3,englman4,englman5,englman6,Sj=0000F6qvist,moore1,moore2,moore3,Kendrick1,Xiao1,izmaylov1,izmaylov2,izmaylov3,gross1,gross2,gross3,agostini2,kinaiak1,Yarkony1,Yarkony2,Yarkony3,kinaiak2}.
However, all of these works related to the geometrical or topological
properties, like Berry phase etc... of natural conical intersections. 

In our earlier papers \cite{LICI1,LICI2} we have calculated the Berry
phase around the light-induced conical intersection formed in the
$\mathrm{Na_{2}}$ molecule in the presence of external electric field.
Due to non vanishing transition dipole moment the $663$ nm light
can resonantly couple the $X^{1}\Sigma_{g}^{+}$ and $A^{1}\Sigma_{u}^{+}$
electronic states giving rise to a light-induced conical intersection.
During the calculations 2$\times$ 2 Floquet form has been assumed
for the Hamiltonian. Combining the calculation of adiabatic-to-diabatic
transformation angle \cite{Domcke1,Baer1,Graham1,Baer2} with the
time-independent line integral technique \cite{Baer2}, we could calculate
the Berry phase for different close contours which either surrounded
or not surrounded the LICI. Obtained results were similar to those
calculated for CIs given in nature demonstrating that a ``true''
CI has been found. 

In the present work we would like to go beyond the time-independent
description. Here we intend to provide calculations for the geometric
phase of the light-induced conical intersection applying the time-dependent
adiabatic approximation proposed by Berry in his famous work\cite{Berry}.
Our showcase example is the $\mathrm{D_{2}^{+}}$ molecule. We demonstrate
that for certain initial conditions by assuming again 2$\times$ 2
Floquet form for the Hamiltonian the time-dependent description can
provide similar results to the time-independent one. 

The article is structured as follows. In the next section, our working
Hamiltonian and the criteria for obtaining LICI are described. The
applied method and the basic formulas, as well as the numerical procedures
are briefly summarized in the third section. In the forth section
the numerical results are presented and discussed. Finally we present
conclusions in Section five.

\section{The Hamiltonian}

Let us define the Hamiltonian which governs the dynamics of the $\mathrm{D_{2}^{+}}$
molecule. Two electronic eigenstates $V_{1}(R)$ (ground, $1s\sigma_{g}$)
and $V_{2}(R)$ (excited, $2p\sigma_{u}$) are included in the Hamiltonian
which are coupled by a running laser wave (see in Figure \ref{potential}).
The non-vanishing transition dipole matrix element $d(R)$$\left(=-\left\langle \Phi_{1}^{e}\left|\sum_{j}r_{j}\right|\Phi_{2}^{e}\right\rangle \right)$
is responsible for the light-induced electronic transition. The corresponding
Born-Oppenheimer potentials and the transition dipole were taken from
\cite{D2,D2a}. As the nuclear coordinate $R$ and the molecular orientation
$\theta$ are taken as parameters during the calculations our Hamiltonian
is defined by the potential energies and the laser-molecule interaction.
This interaction is given in the dipole approximation as the scalar
product of the transition dipole moment $\overrightarrow{d}$ and
the electric field vector $\overrightarrow{\varepsilon}$:

\begin{equation}
\overrightarrow{d}\cdot\overrightarrow{\varepsilon}=\epsilon_{0}d(R)\cos\theta cos\left(\omega_{L}t\right).\label{eq:light-matter}
\end{equation}
In Eq. (\ref{eq:light-matter}) $\epsilon_{0}$ is the maximum laser
field amplitude, $I_{0}$ $(\sim\epsilon_{0}^{2})$ is the laser intensity,
$\theta$ denotes the angle between the polarization direction and
the direction of the transition dipole $d(R)$ and $\omega_{L}$ is
the laser frequency which couples the two electronic states at $R=5$
a.u. nuclear distance ($\omega_{L}=1.359\,eV$). 

Let us represent the Hamiltonian in the Floquet picture. Therefore,
the original Hamiltonian is transformed into an equivalent static
problem by using the leading term in the Fourier series expansion
of the solution of the time-dependent Schrödinger equation. Then the
field-dressed form reads

\begin{eqnarray}
\mathbf{\hat{H}} & = & \left(\begin{array}{cc}
V_{1}(R) & (\epsilon_{0}/2)d(R)\cos\theta\\
(\epsilon_{0}/2)d(R)\cos\theta & V_{2}(R)-\hbar\omega_{L}
\end{array}\right)\label{eq: Hamiltonian}
\end{eqnarray}
In this dressed state representation the interaction between the molecule
and the electromagnetic field is obtained by shifting the energy of
the excited potential curve by $\hbar\omega_{L}$. This picture is
often used to explain various phenomena in the area of strong field
physics whenever only net one-photon is absorbed by the molecule. 

As a results of the dressed state representation a crossing is formed
between the diabatic ground and the dressed excited potential energy
curves. After diagonalizing the diabatic potential matrix Eq. \ref{eq: Hamiltonian},
the resulting adiabatic or light-induced surfaces ($V_{lower}$ and
$V_{upper}$) form a light-induced conical intersection (see in Figure
\ref{potential}) whenever the following conditions are fulfilled\cite{Lenz2,LICI1}:
\begin{equation}
cos\theta=0\quad(\theta=\frac{\pi}{2})\qquad\mathrm{and}\qquad V_{1}(R)=V_{2}(R)-\hbar\omega_{L}.\label{criteria}
\end{equation}

An important feature of the light-induced conical intersections as
compared to the natural CIs is that their fundamental characteristics
can be modified by the external field. It has already been shown that
the intensity of the field determines the strength of the nonadiabatic
coupling, namely the steepness of the cone, while the energy of the
field specifies the position of the LICI.

\section{The methodology and the numerical details }

The main subject of this section is to obtain the appropriate expression
so as to compute the geometric phase. 

Let us consider again the working Hamiltonian Eq. \ref{eq: Hamiltonian}
which is parametrized by $R$ and $\theta$. If the system starts
in an eigenstate $\Phi(R,\theta)$ with an energy $E(R,\theta)$,
then it evolves into the state $\exp[-iE(R,\theta)t]\Phi(R,\theta)$.
Now let the parameters vary slowly, $R=R(t)$ and $\theta=\theta(t)$
\footnote{Here we allow a very slow time dependence of the $R$ and $\theta$
parameters, so as to assume an implicit adiabatic time dependence
of the working Floquet Hamiltonian. }, then due to the adiabatic theorem, the eigenstates $\Phi(R,\theta)$
are replaced by one of the actual eigenstates $\Phi(R(t),\theta(t))$.
If both $R\left(t\right)$ and $\theta\left(t\right)$ are periodic
functions of time with a period of $T$ they describe a closed path
in the configuration space. That is, for the time $t=T$ the initial
state $\Phi(R(t=0),\theta(t=0))$ evolves into the final state which
is identical with the initial state except for a phase factor: $\Phi(R(T),\theta(T))=\exp\left(i\chi\right)\cdot\Phi\left(R(0),\theta(0)\right).$
It is easy to see that the phase factor is identical with the autocorrelation
function $\left(C\left(t\right)=\left\langle \Phi(R(0),\theta(0))|\Phi(R(t),\theta(t))\right\rangle \right)$
at time $T$. Berry showed \cite{Berry} that $\chi$ is the sum of
$\delta=-\int_{0}^{T}E(R(t'),\theta(t'))dt'$ and a quantity $\gamma$,
latter is called the adiabatic phase. Here $\chi$ and $\delta$ are
the overall and the dynamical phases, respectively. 

Both the $\chi$ and the $\delta$ functions can be generalized for
any arbitrary time $t$. To obtain the actual expressions for the
$\chi(t)$ and $\delta(t)$ phases we refer to the work of Mukunda
and Simon \cite{Mukunda}. Among others they have pointed out that
the overall phase is the argument of the autocorrelation function

\begin{equation}
\chi(t)=arg\left\langle \Phi(R(0),\theta(0))|\Phi(R(t),\theta(t))\right\rangle \label{eq:autocorrelation}
\end{equation}
 and the dynamical phase is as follows

\begin{equation}
\delta(t)=i\int_{0}^{t}\left\langle \Phi(R(t'),\theta(t'))|\dot{\Phi}(R(t'),\theta(t'))\right\rangle dt'.\label{eq:dynamicalphase1}
\end{equation}

Aharonov and Anandan \cite{Aharonov} pointed out that $\gamma$ is
a purely geometrical property of the path which is parametrically
defined by the functions $R=R(t)$ and $\theta=\theta(t)$ and can
be calculated as the difference of the $\chi(t)$ and $\delta(t)$
at the end of the closed path. Therefore its value depends only upon
the contour followed by the system in the configuration space. Hence
the name of $\gamma$ is geometric phase \footnote{From now on we will refer to $\gamma$ as Berry, geometric or adiabatic
phase. }. If $\Phi(R(t),\theta(t))$ is the solution of the dynamical Schrödinger
equation, i.e., satisfies the $i\hbar\left|\dot{\Phi}(R(t),\theta(t))\right\rangle =\hat{H}(R(t),\theta(t))\left|\Phi(R(t),\theta(t))\right\rangle $,
then we obtain for Eq. \ref{eq:dynamicalphase1}

\begin{equation}
\delta(t)=\frac{1}{\hbar}\int_{0}^{t}\left\langle \Phi(R(t'),\theta(t'))|\hat{H}(R(t),\theta(t))|\Phi(R(t'),\theta(t'))\right\rangle dt'.\label{eq:dinamicalphase2}
\end{equation}

Using Eqs. \ref{eq:autocorrelation} and \ref{eq:dinamicalphase2}
one can calculate the geometric phase $\gamma$, as difference of
the $\chi(t)$ and $\delta(t)$ expressions at the end of the closed
path for adiabatically slow changes of the parameters $R\left(t\right)$
and $\theta\left(t\right)$ over the whole path. As we do not know
in advance how slow change can be considered as an adiabatic one during
the numerical simulations we consider the quantity
\begin{equation}
\widetilde{\gamma}=\chi\left(T\right)-\delta\left(T\right)\label{eq:gamma-approx}
\end{equation}
 as an approximation for the Berry phase for the given contour.

To get the $\Phi(R(t),\theta(t))$ wave function we have solved numerically
the time-dependent Schrödinger equation by using implicit 4th order
Runge-Kutta integrator with Gaussian points\cite{rk4imp} implemented
in the GNU Scientific Library\cite{GSL-1.13}.

\section{Results and discussion}

So as to understand the meaning of the numerical results to be presented
in this paper, we first discuss the geometrical situation for which
the above approach is applied and then analyze the numerical results. 

\begin{figure}[p]
\begin{centering}
\includegraphics[width=0.5\textwidth]{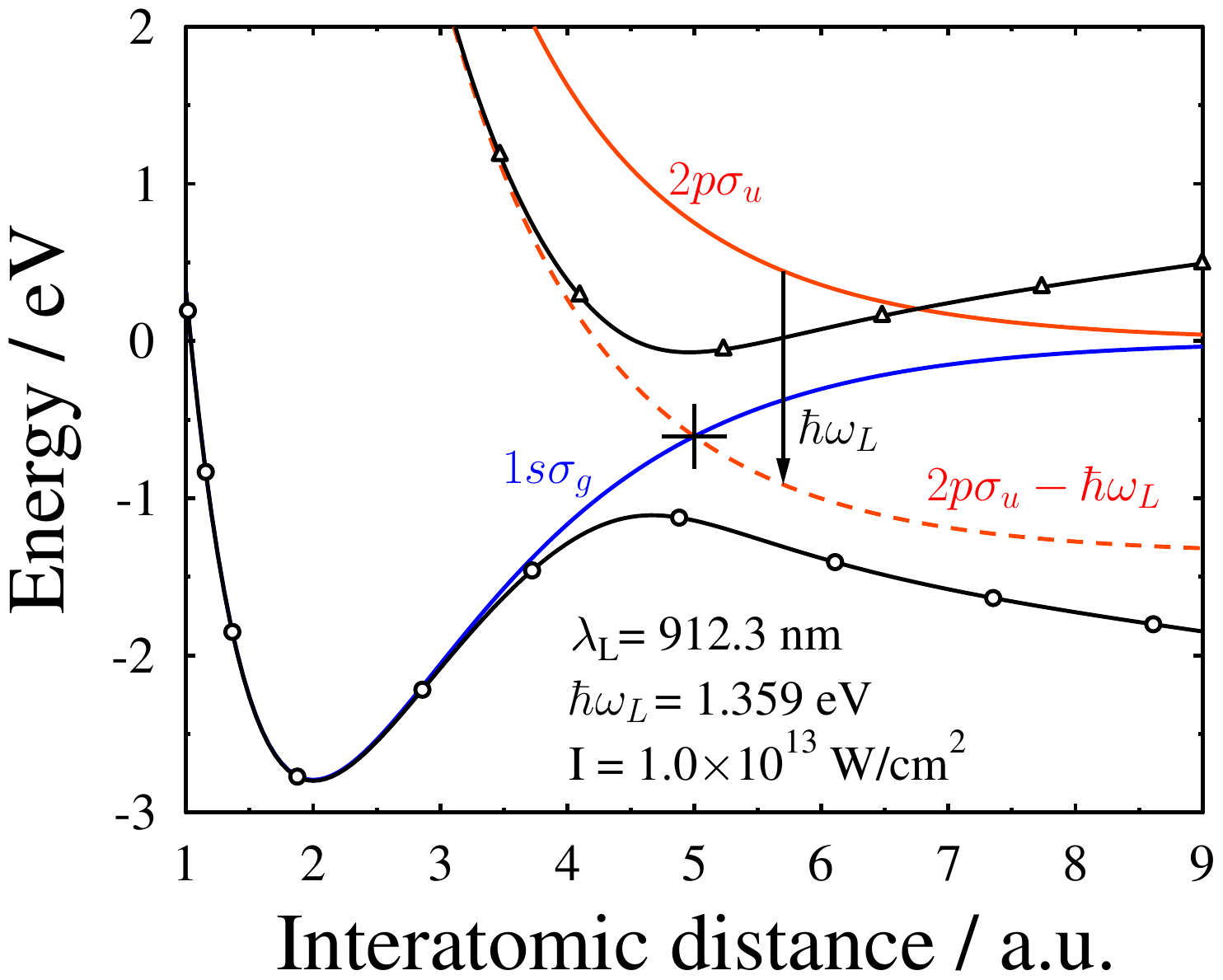}
\par\end{centering}

\caption{\label{potential}A cut through the potential energy surface of the
$\mathrm{D}_{2}^{+}$ molecule as a function of interatomic separation.
Diabatic energies of the ground $V_{1}(R)$ $\left(1s\sigma_{g}\right)$
and the first excited $V_{2}(R)$ $\left(2p\sigma_{u}\right)$ states
are displayed with solid blue and red lines, respectively. The field
dressed excited state ($2p\sigma_{u}-\hbar\omega_{L}$ ; dashed red
line) forms a light induced conical intersection (LICI) with the ground
state. For the case of a laser frequency $\hbar\omega_{L}=1.359eV$
and field intensity of $1\times10^{13}\frac{W}{cm^{2}}$ a cut through
the adiabatic surfaces at $\theta=0$ (parallel to the field) is also
shown by solid black lines marked with circles ($V_{lower}$) and
triangles ($V_{upper}$). We denote with a cross the position of the
LICI ($R_{LICI}=5.00\,a.u.$).}
\end{figure}

\begin{figure}[p]
\begin{centering}
\includegraphics[width=0.5\textwidth]{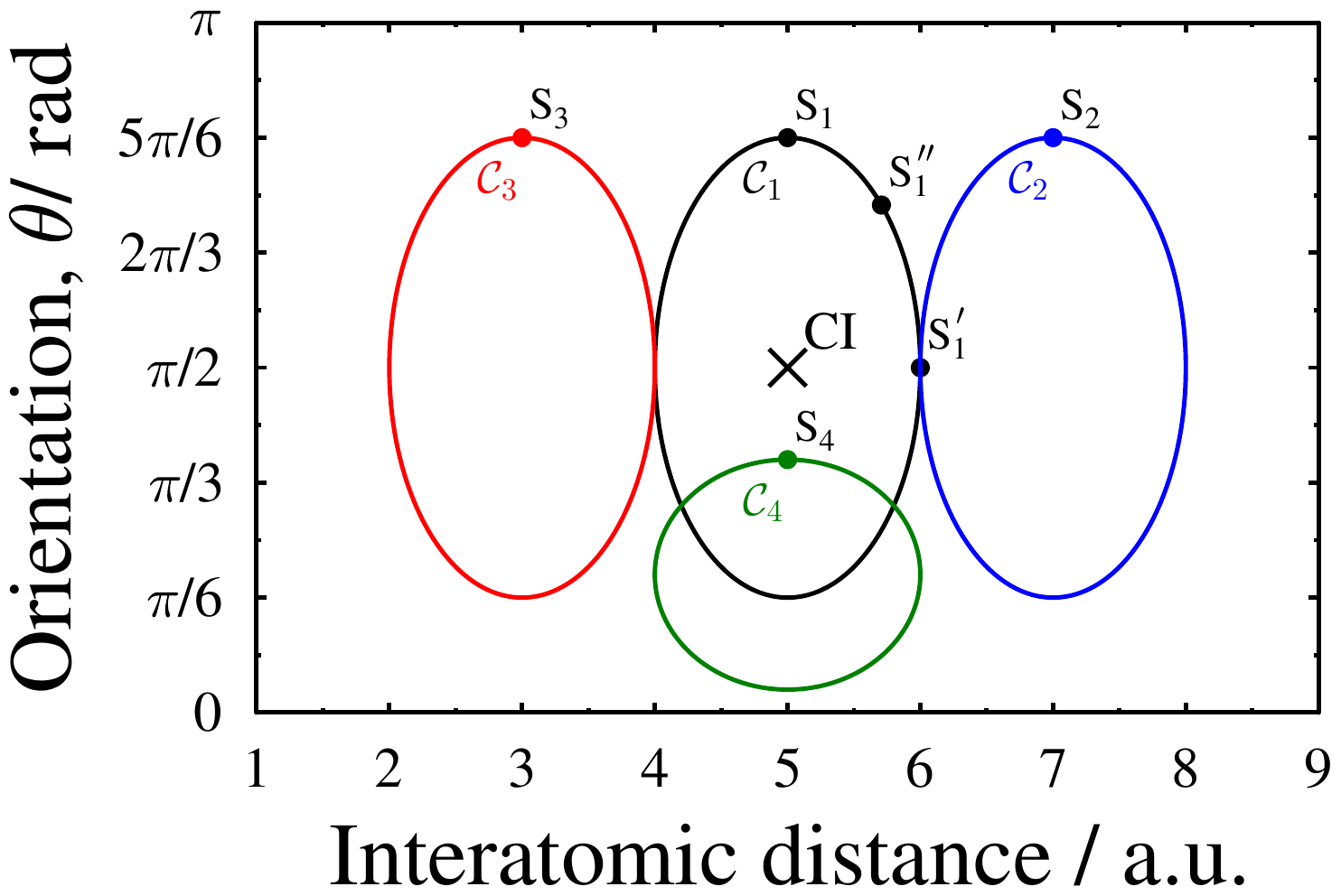}
\par\end{centering}

\caption{\label{geometry}Geometrical arrangement of the contours used in the
geometrical phase calculations. Four different geometrical arrangements
are applied and only one curve surrounds the LICI. The black cross
shows the position of the LICI. Dots denote the starting points of
the different simulations in the configuration space.}
\end{figure}

The light-induced adiabatic states ($V_{lower}$ and $V_{upper}$)
as well as the position of the corresponding light-induced conical
intersection are displayed in Figure \ref{potential}. The geometrical
arrangement used as contours in the geometric phase calculations are
shown in Figure \ref{geometry}. Here, four different ellipses are
presented but only one of them surrounds the light-induced conical
intersection. The numerical calculations take place along these closed
paths characterized by their centers $(R_{c},\theta_{c})$ and radii
$\left(\rho_{R},\rho_{\theta}\right)$ . The actual position is given
by an angle $\beta(t)=\beta_{0}+t/T\cdot2\pi$:
\begin{eqnarray}
R\left(t\right) & = & R_{c}+\rho_{R}\cos\beta\left(t\right)\label{position}\\
\theta\left(t\right) & = & \theta_{c}-\rho_{\theta}\sin\beta\left(t\right).\nonumber 
\end{eqnarray}
The applied parameters for the different contours are displayed in
Table \ref{tab:Parameters}. In Table \ref{tab:gamma} the obtained
approximated values ($\widetilde{\gamma}$) for the geometric phase
$\gamma$ are presented (in unit of $\pi$) for the contour $\mathcal{C}_{1}$
which encircles the LICI with the initial wave function chosen to
be the lower lying eigenstate of the Hamiltonian Eq. \ref{eq: Hamiltonian}
at point $S_{1}$. The approximation is based on the difference of
the argument of the autocorrelation function (Eq. \ref{eq:autocorrelation})
and the dynamical phase (Eq. \ref{eq:dinamicalphase2}) at the end
of the path (see Eq. \ref{eq:gamma-approx}). The applied photon energy
is $\hbar\omega_{L}=1.359\,eV$. In the rows of Table \ref{tab:gamma}
the different field intensities are chosen to be between $I=1\times10^{10}\frac{W}{cm^{2}}$
and $I=1\times10^{15}\frac{W}{cm^{2}}$. In the columns, the periodic
times of the round transport of the ellipse are indicated as a unit
of the periodic time of the laser pulse ($\frac{2\pi}{\omega_{L}}$).
The larger the value of $T$ indicated here, the more adiabatic the
process of encircling the ellipse. As the contour $\mathcal{C}_{1}$
surrounds only a single conical intersection the value of the geometric
phase $\gamma$ is expected to be $\pm\pi$ and so for long enough
T values $\widetilde{\gamma}$ should be in the close vicinity of
$(2n+1)\pi$ (where $n$ in an integer). We can observe in Table \ref{tab:gamma}
that except for the lowest studied intensities the value of $\widetilde{\gamma}$
is really close to $\pi$, whenever $T\geq500$$\times$$\frac{2\pi}{\omega_{L}}$.
The numerical problem at small intensities are related to the fact
that for the field free case (zero intensity) the value of $\gamma$
should be zero. As a consequence, in weak fields we need extremely
slow surrounding of the contour to be able to consider it as an adiabatic
one. For the extremely large values of T question arises about the
accuracy concerning the numerical integration of the Schrödinger equation.
As a simple check for this issue we have also performed numerical
integration over the same contour for the field free case. The obtained
$\widetilde{\gamma}$ values are displayed in the first row of Table
\ref{tab:gamma}. All of these values are close to the expected value
of $\gamma=0$.

For larger intensities the adiabatic region can be reached before
$T=500\times\frac{2\pi}{\omega_{L}}$. Table \ref{tab:gamma} shows
that for intensities larger than $I=1\times10^{13}\frac{W}{cm^{2}}$
the beginning of this adiabatic region is moving towards larger values
of T with the increasing intensities. This effect is related to the
fact that at higher intensities the derivative of the adiabatic potentials
($V_{lower}$ and $V_{upper}$) respect to the position on the contour
(controlled by parameter $\beta\left(t\right)$) become significantly
larger than the same derivatives of the diabatic ones. As a result,
slower change is requested in the value of the $\beta\left(t\right)$
parameter so as to consider the process being adiabatic.

\begin{figure}[p]
\begin{centering}
\includegraphics[width=0.5\textwidth]{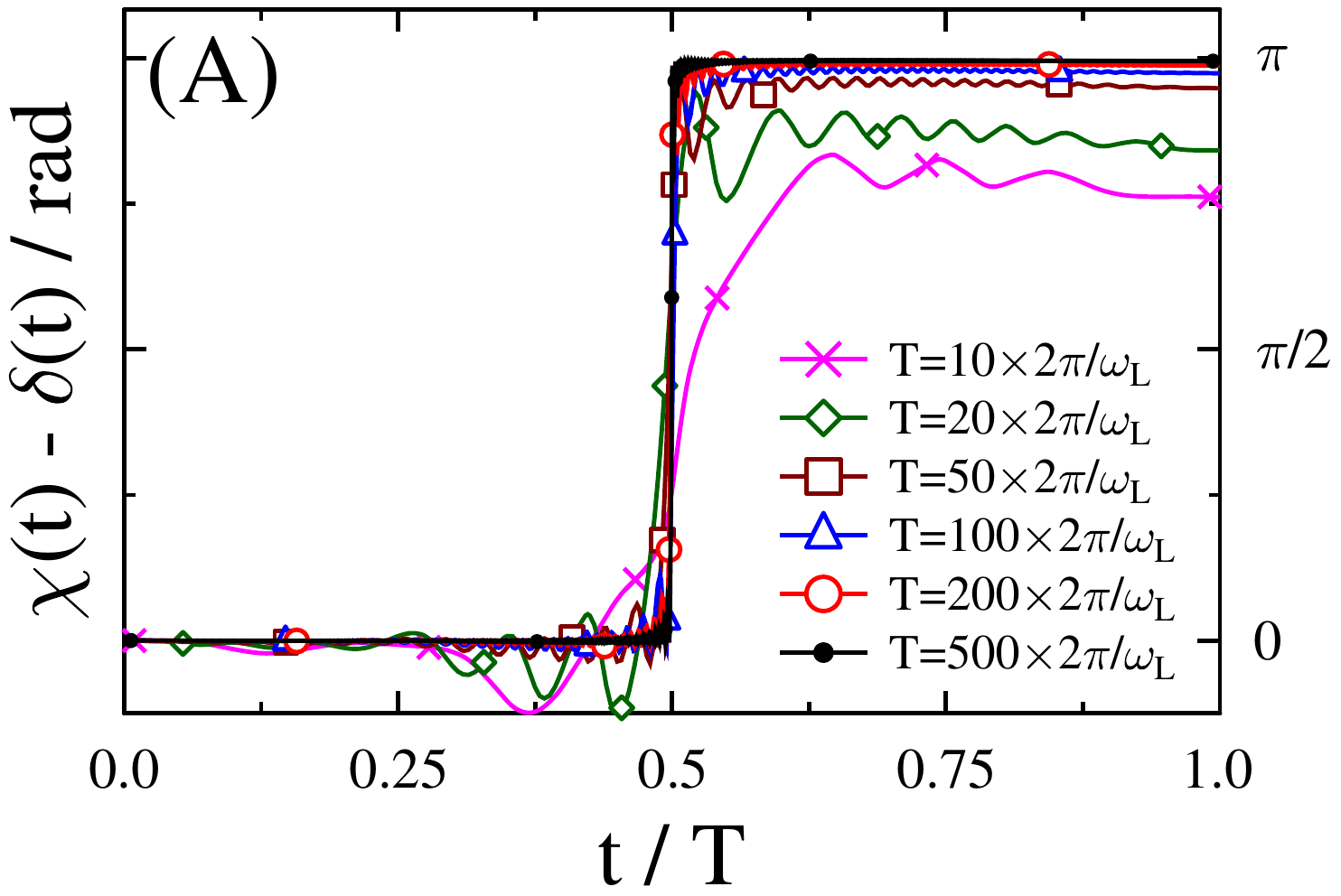} 
\par\end{centering}

\begin{centering}
\includegraphics[width=0.5\textwidth]{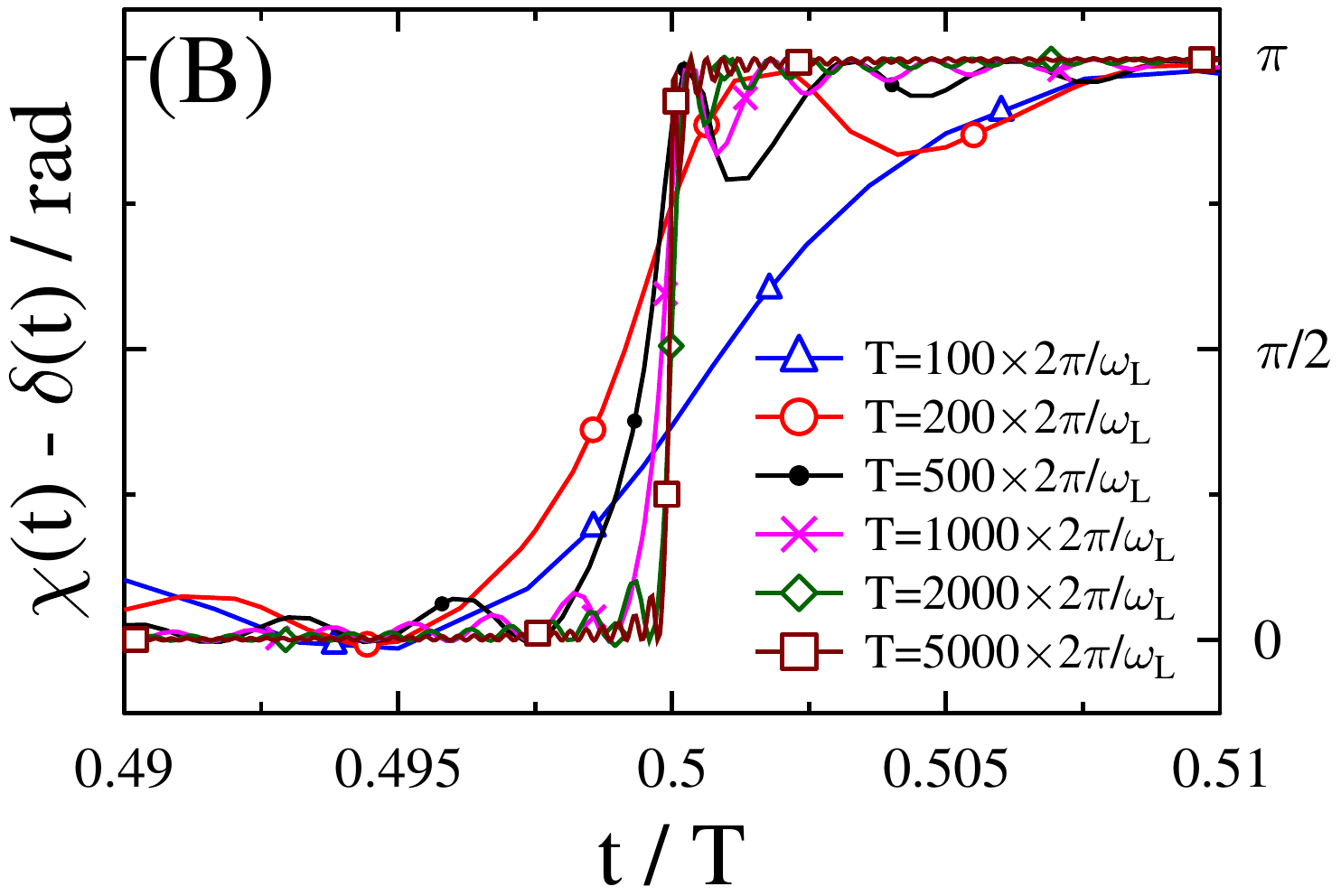} 
\par\end{centering}

\begin{centering}
\includegraphics[width=0.5\textwidth]{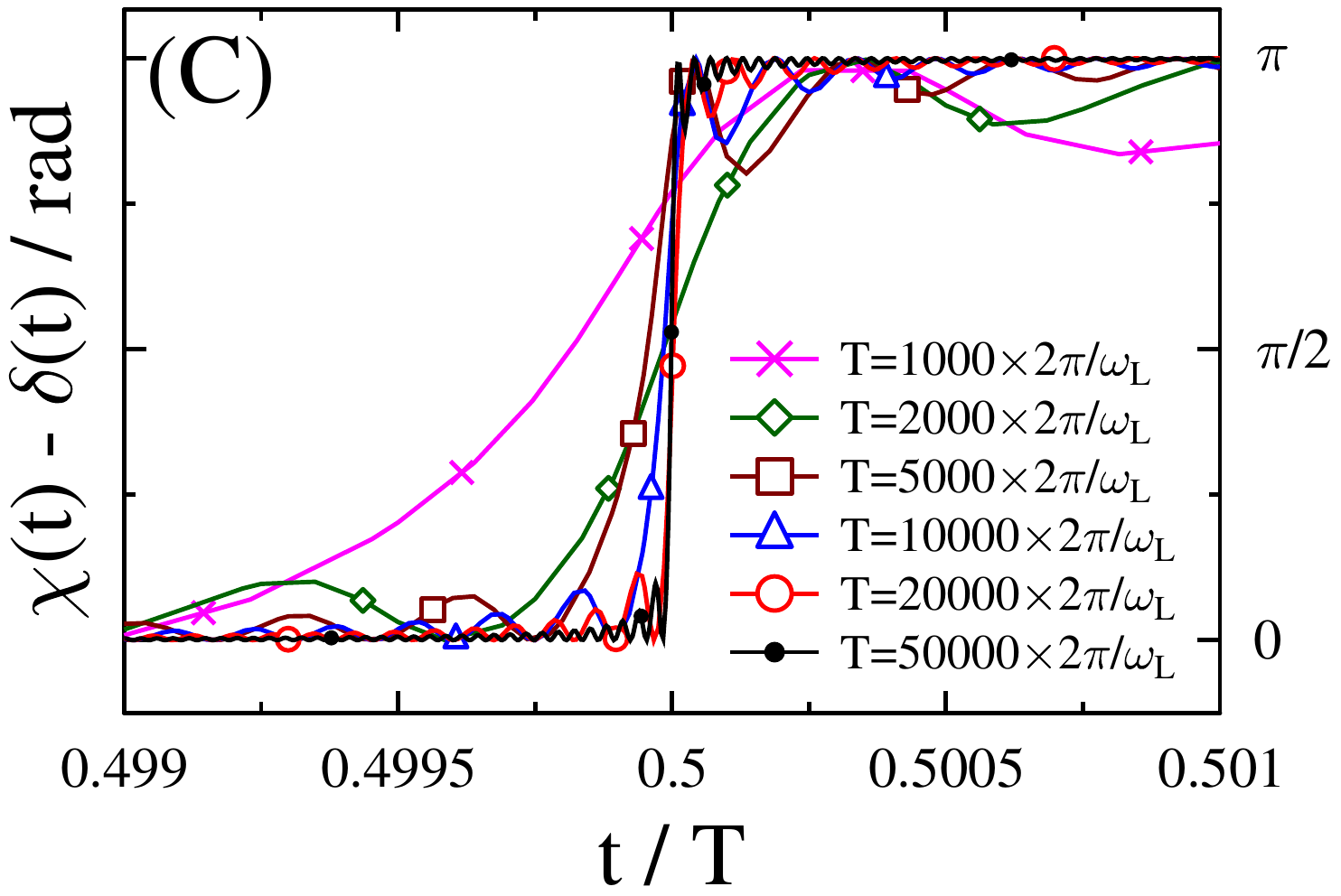} 
\par\end{centering}

\caption{\label{gammak} The difference of the total and the dynamical phases
as a function of time. This quantity provides the value of the geometric
phase $\gamma$ at $t/T=1$ if the $T$ is long enough. $T$ is the
periodic time encircling the ellipse. The computation starts at the
point $S_{1}$ of the ellipse which surrounds the LICI (see on Fig.\ref{geometry}).
Different time intervals are depicted. In panel (A) the whole time
period $t/T\left\{ 0,...1\right\} $, in panel (B) the $t/T\left\{ 0.49,...0.51\right\} $
and in panel (C) the $t/T\left\{ 0.499...0.501\right\} $ are figured.
The applied intensity is $I=1\times10^{13}\frac{W}{cm^{2}}$. }
 
\end{figure}

In Figure \ref{gammak} the difference of the $\chi(t)$ and $\delta(t)$
functions are displayed as a function of time with three different
time resolutions. Results are presented of the set of simulations
for which the $\widetilde{\gamma}$ values are displayed in Table
\ref{tab:gamma} at $I=1\times10^{13}\frac{W}{cm^{2}}$ field intensity.
It can be seen that the different curves display different shapes
at $t/T=0.5$, but all of them possess relatively sudden jumps ranging
from near zero to close to the final value. The phase jumps always
take place at that position of $t/T$, where the value of the autocorrelation
function tends to zero. In the panels of Figure \ref{gammak} the
positions of the phase jumps are always positioned at $t/T=0.5$.
But this happens due to symmetry reasons. In the current geometrical
arrangement (see on Figure \ref{geometry}), the center of the ellipse
is also the position of the LICI. In this situation, four symmetry
points can be considered concerning the starting position of the encircling
of the ellipse. These symmetry points are the endpoints of the small
and the large axes of the ellipse. If one starts to circle the ellipse
at one of these points the phase jumps always occur at $t/T=0.5$.
If this process starts from a different point then the phase jump
happens at another value of $t/T$. E.g. for the contour $\mathcal{C}_{1}$
the phase jump occurs at $t/T=0.5$ if the starting points are $S_{1}$
or $S_{1}^{\prime}$ but for starting point $S_{1}^{\prime\prime}$
the jump happens at $t/T\cong0.44$. Nevertheless, the values of the
phase jumps are always very close to an odd multiple of $\pi$. The
longer the encircling time, the steeper the phase jump, but its value
is getting closer and closer to the final value of $\widetilde{\gamma}$.
This effect is clearly recognizable as long as the time resolution
gets finer (see on panels of Figure \ref{gammak}).

We have also computed the value of the approximate geometric phase
$\widetilde{\gamma}$ along those ellipses which do not surround LICI.
Obtained results are always very close to zero. 

For completeness we have performed similar calculations on the upper
adiabatic surfaces as well. All of these calculations provide the
same results as for the case of lower surface but always with an opposite
sing for $\widetilde{\gamma}$. Table \ref{tab:gamma2} displays some
values of $\widetilde{\gamma}$ at $I=1\times10^{13}\frac{W}{cm^{2}}$
field intensity. We notice that for contour $\mathcal{C}_{1}$ starting
the simulation at point $S_{1}^{\prime}$ the calculated values of
$\widetilde{\gamma}$ are always around of $\pm3\pi$ or $\pm5\pi$
depending upon the actual speed of the surrounding. (All of these
values are odd multiples of $\pi$ so they are in agreement with the
expected value of the geometric phase $\gamma=\pi$.) This uncertainty
is clearly related to the fact that the correct value of the autocorrelation
function is zero at $t/T=0.5$ and therefore it is extremely hard
to follow its argument during the numerical simulations. 

\begin{table}[p]
\begin{tabular}{|c||c|c|c|c|c|}
\hline 
Contour & $R_{c}\left[a.u.\right]$ & $R_{\theta}\left[rad\right]$ & $\rho_{c}\left[a.u.\right]$ & $\rho_{\theta}\left[rad\right]$ & $\beta_{0}\left[rad\right]$\tabularnewline
\hline 
\hline 
$\mathcal{C}_{1}$ & 5 & $\pi/2$ & 1 & $\pi/3$ & $-\pi/2$ for $S_{1}$;\quad{}0 for $S_{1}^{\prime}$;\quad{}$-\pi/4$
for $S_{1}^{\prime\prime}$\tabularnewline
\hline 
$\mathcal{C}_{2}$ & 7 & $\pi/2$ & 1 & $\pi/3$ & $-\pi/2$ for $S_{2}$\tabularnewline
\hline 
$\mathcal{C}_{3}$ & 3 & $\pi/2$ & 1 & $\pi/3$ & $-\pi/2$ for $S_{3}$\tabularnewline
\hline 
$\mathcal{C}_{4}$ & 5 & $\pi/5$ & 1 & $\pi/6$ & $-\pi/2$ for $S_{4}$\tabularnewline
\hline 
\end{tabular}

\caption{\label{tab:Parameters}Parameters of the applied contours in the configuration
space corresponding to Figure \ref{geometry} and Eq. \ref{position}.}
\end{table}
\begin{table}[p]
\begin{tabular}{|c||c|c|c|c|c|c|}
\hline 
Intensity & \multicolumn{6}{c|}{$T\;\;\left[2\pi/\omega_{L}\right]$}\tabularnewline
$\left[W/cm^{2}\right]$ & \multicolumn{1}{c}{10} & \multicolumn{1}{c}{20} & \multicolumn{1}{c}{50} & \multicolumn{1}{c}{100} & \multicolumn{1}{c}{200} & 500\tabularnewline
\hline 
\hline 
field free & \texttt{<1$\cdot10^{-6}$} & \texttt{<1$\cdot10^{-6}$} & \texttt{<1$\cdot10^{-6}$} & \texttt{<1$\cdot10^{-6}$} & \texttt{<1$\cdot10^{-6}$} & \texttt{<1$\cdot10^{-6}$}\tabularnewline
\hline 
$1\cdot10^{10}$  & \texttt{0.4465 } & \texttt{-8.2194 } & \texttt{0.0682 } & \texttt{13.7408 } & \texttt{-75.0925 } & \texttt{32.9789 }\tabularnewline
\hline 
$3\cdot10^{10}$ & \texttt{0.0663 } & \texttt{-7.7439 } & \texttt{-3.7613 } & \texttt{8.8010 } & \texttt{-29.9996 } & \texttt{-29.6432 }\tabularnewline
\hline 
$5\cdot10^{10}$ & \texttt{-0.1732 } & \texttt{-7.4972 } & \texttt{-5.1357 } & \texttt{4.5659 } & \texttt{-24.4448 } & \texttt{-96.5521 }\tabularnewline
\hline 
$1\cdot10^{11}$ & \texttt{-0.5458 } & \texttt{-7.2619 } & \texttt{-3.5758 } & \texttt{7.7951 } & \texttt{-51.2940 } & \texttt{-17.0145 }\tabularnewline
\hline 
$3\cdot10^{11}$ & \texttt{-0.9243 } & \texttt{2.2194 } & \texttt{-9.2701 } & \texttt{-1.6251 } & \texttt{-4.4822 } & \texttt{-0.4876 }\tabularnewline
\hline 
$5\cdot10^{11}$ & \texttt{-0.6192 } & \texttt{1.9429 } & \texttt{-0.6340 } & \texttt{-3.7004 } & \texttt{-0.4275 } & \texttt{0.4730 }\tabularnewline
\hline 
$1\cdot10^{12}$ & \texttt{1.1093 } & \texttt{-4.9457 } & \texttt{-1.3928 } & \texttt{0.2718 } & \texttt{0.6346 } & \texttt{0.8578 }\tabularnewline
\hline 
$3\cdot10^{12}$ & \texttt{0.3088 } & \texttt{0.3741 } & \texttt{0.7764 } & \texttt{0.8911 } & \texttt{0.9463 } & \texttt{0.9785 }\tabularnewline
\hline 
$5\cdot10^{12}$ & \texttt{0.4448 } & \texttt{0.7695 } & \texttt{0.8962 } & \texttt{0.9479 } & \texttt{0.9741 } & \texttt{0.9896 }\tabularnewline
\hline 
$1\cdot10^{13}$ & \texttt{0.7615 } & \texttt{0.8417 } & \texttt{0.9481 } & \texttt{0.9741 } & \texttt{0.9871 } & \texttt{0.9948 }\tabularnewline
\hline 
$3\cdot10^{13}$ & \texttt{5.4996 } & \texttt{2.6332 } & \texttt{0.9582 } & \texttt{0.9795 } & \texttt{0.9898 } & \texttt{0.9959 }\tabularnewline
\hline 
$5\cdot10^{13}$ & \texttt{7.7773 } & \texttt{11.9147 } & \texttt{2.9448 } & \texttt{0.9765 } & \texttt{0.9883 } & \texttt{0.9953 }\tabularnewline
\hline 
$1\cdot10^{14}$ & \texttt{8.1765 } & \texttt{21.4233 } & \texttt{14.7622 } & \texttt{0.9681 } & \texttt{0.9846 } & \texttt{0.9939 }\tabularnewline
\hline 
$3\cdot10^{14}$ & \texttt{9.4209 } & \texttt{35.7743 } & \texttt{104.6614 } & \texttt{54.7573 } & \texttt{2.9740 } & \texttt{0.9897 }\tabularnewline
\hline 
$5\cdot10^{14}$ & \texttt{5.2325 } & \texttt{35.7902 } & \texttt{146.1297 } & \texttt{199.6102 } & \texttt{20.9600 } & \texttt{0.9868 }\tabularnewline
\hline 
$1\cdot10^{15}$ & \texttt{0.6056 } & \texttt{37.9231 } & \texttt{190.5541 } & \texttt{418.5138 } & \texttt{252.6389 } & \texttt{0.9814 }\tabularnewline
\hline 
\end{tabular}\\
\begin{tabular}{|c||c|c|c|c|c|c|}
\hline 
Intensity & \multicolumn{6}{c|}{$T\;\;\left[2\pi/\omega_{L}\right]$}\tabularnewline
$\left[W/cm^{2}\right]$ & \multicolumn{1}{c}{1000} & \multicolumn{1}{c}{2000} & \multicolumn{1}{c}{5000} & \multicolumn{1}{c}{10000} & \multicolumn{1}{c}{20000} & 50000\tabularnewline
\hline 
\hline 
field free & \texttt{1$\cdot10^{-6}$} & \texttt{2$\cdot10^{-6}$} & \texttt{4$\cdot10^{-6}$} & \texttt{8$\cdot10^{-6}$} & \texttt{15$\cdot10^{-6}$} & \texttt{24$\cdot10^{-6}$}\tabularnewline
\hline 
$1\cdot10^{10}$ & \texttt{-305.7453 } & \texttt{-82.5002 } & \texttt{-192.7763 } & \texttt{-60.5332 } & \texttt{-29.1699 } & \texttt{-10.6957 }\tabularnewline
\hline 
$3\cdot10^{10}$ & \texttt{-148.5720 } & \texttt{-49.4776 } & \texttt{-12.8597 } & \texttt{-5.5952 } & \texttt{-2.2615 } & \texttt{-0.3007 }\tabularnewline
\hline 
$5\cdot10^{10}$ & \texttt{-39.6079 } & \texttt{-13.1177 } & \texttt{-3.8062 } & \texttt{-1.3639 } & \texttt{-0.1775 } & \texttt{0.5295 }\tabularnewline
\hline 
$1\cdot10^{11}$ & \texttt{-5.7173 } & \texttt{-2.0801 } & \texttt{-0.1961 } & \texttt{0.4041 } & \texttt{0.7023 } & \texttt{0.8809 }\tabularnewline
\hline 
$3\cdot10^{11}$ & \texttt{0.3002 } & \texttt{0.6534 } & \texttt{0.8618 } & \texttt{0.9309 } & \texttt{0.9655 } & \texttt{0.9862 }\tabularnewline
\hline 
$5\cdot10^{11}$ & \texttt{0.7400 } & \texttt{0.8705 } & \texttt{0.9483 } & \texttt{0.9741 } & \texttt{0.9870 } & \texttt{0.9948 }\tabularnewline
\hline 
$1\cdot10^{12}$  & \texttt{0.9293 } & \texttt{0.9646 } & \texttt{0.9859 } & \texttt{0.9929 } & \texttt{0.9964 } & \texttt{0.9986 }\tabularnewline
\hline 
$3\cdot10^{12}$  & \texttt{0.9893 } & \texttt{0.9946 } & \texttt{0.9978 } & \texttt{0.9989 } & \texttt{0.9994 } & \texttt{0.9998 }\tabularnewline
\hline 
$5\cdot10^{12}$ & \texttt{0.9948 } & \texttt{0.9974 } & \texttt{0.9990 } & \texttt{0.9995 } & \texttt{0.9997 } & \texttt{0.9999 }\tabularnewline
\hline 
$1\cdot10^{13}$ & \texttt{0.9974 } & \texttt{0.9987 } & \texttt{0.9995 } & \texttt{0.9997 } & \texttt{0.9999 } & \texttt{0.9999 }\tabularnewline
\hline 
$3\cdot10^{13}$ & \texttt{0.9980 } & \texttt{0.9990 } & \texttt{0.9996 } & \texttt{0.9998 } & \texttt{0.9999 } & \texttt{1.0000 }\tabularnewline
\hline 
$5\cdot10^{13}$ & \texttt{0.9977 } & \texttt{0.9988 } & \texttt{0.9995 } & \texttt{0.9998 } & \texttt{0.9999 } & \texttt{1.0000 }\tabularnewline
\hline 
$1\cdot10^{14}$ & \texttt{0.9969 } & \texttt{0.9985 } & \texttt{0.9994 } & \texttt{0.9997 } & \texttt{0.9998 } & \texttt{1.0000 }\tabularnewline
\hline 
$3\cdot10^{14}$ & \texttt{0.9949 } & \texttt{0.9974 } & \texttt{0.9990 } & \texttt{0.9995 } & \texttt{0.9997 } & \texttt{1.0000 }\tabularnewline
\hline 
$5\cdot10^{14}$ & \texttt{0.9934 } & \texttt{0.9967 } & \texttt{0.9987 } & \texttt{0.9993 } & \texttt{0.9997 } & \texttt{1.0001 }\tabularnewline
\hline 
$1\cdot10^{15}$ & \texttt{0.9907 } & \texttt{0.9954 } & \texttt{0.9981 } & \texttt{0.9990 } & \texttt{0.9995 } & \texttt{1.0003 }\tabularnewline
\hline 
\end{tabular}

\caption{\label{tab:gamma}The difference of the total and dynamical phases
in the units of $\pi$ at the end of the paths $\left(t=T\right)$
$\mathcal{C}_{1}$ surrounding the LICI (see on Fig.\ref{geometry})
The staring point of the surrounding is $S_{1}$ and the initial wave
function is chosen to be on the lower adiabatic surface.}
\end{table}

\begin{table}[p]
\begin{tabular}{|c||c|c|c|c|c|c|c|}
\hline 
Contour/ & \multicolumn{7}{c|}{$T\;\;\left[2\pi/\omega_{L}\right]$}\tabularnewline
Surface & \multicolumn{1}{c}{500} & \multicolumn{1}{c}{1000} & \multicolumn{1}{c}{2000} & \multicolumn{1}{c}{5000} & \multicolumn{1}{c}{10000} & \multicolumn{1}{c}{20000} & 50000\tabularnewline
\hline 
\hline 
$\mathcal{C}_{1}$-$S_{1}$/ lower & \texttt{+0.99483} & \texttt{+0.99741} & \texttt{+0.99871} & \texttt{+0.99948} & \texttt{+0.99973} & \texttt{+0.99985} & \texttt{+0.99993}\tabularnewline
\hline 
$\mathcal{C}_{1}$-$S_{1}$/ upper & \texttt{-0.99483} & \texttt{-0.99742} & \texttt{-0.99871} & \texttt{-0.99948} & \texttt{-0.99974} & \texttt{-0.99987} & \texttt{-0.99995}\tabularnewline
\hline 
$\mathcal{C}_{1}$-$S_{1}^{\prime}$/ lower & \texttt{+4.98611} & \texttt{+2.99305} & \texttt{+2.99652} & \texttt{+4.99859} & \texttt{+4.99926} & \texttt{+2.99957} & \texttt{+4.99965}\tabularnewline
\hline 
$\mathcal{C}_{1}$-$S_{1}^{\prime}$/ upper & \texttt{-4.98611} & \texttt{-2.99306} & \texttt{-2.99653} & \texttt{-4.99861} & \texttt{-4.99931} & \texttt{-2.99966} & \texttt{-4.99987}\tabularnewline
\hline 
$\mathcal{C}_{1}$-$S_{1}^{\prime\prime}$/ lower & \texttt{+0.99483} & \texttt{+0.99741} & \texttt{+0.99871} & \texttt{+0.99948} & \texttt{+0.99973} & \texttt{+0.99985} & \texttt{+0.99993}\tabularnewline
\hline 
$\mathcal{C}_{1}$-$S_{1}^{\prime\prime}$/ upper & \texttt{-0.99483} & \texttt{-0.99742} & \texttt{-0.99871} & \texttt{-0.99948} & \texttt{-0.99974} & \texttt{-0.99987} & \texttt{-0.99995}\tabularnewline
\hline 
$\mathcal{C}_{2}$-$S_{2}$/ lower & \texttt{-0.00185} & \texttt{-0.00093} & \texttt{-0.00047} & \texttt{-0.00019} & \texttt{-0.00011} & \texttt{-0.00007} & \texttt{-0.00002}\tabularnewline
\hline 
$\mathcal{C}_{2}$-$S_{2}$/ upper & \texttt{+0.00185} & \texttt{+0.00092} & \texttt{+0.00046} & \texttt{+0.00019} & \texttt{+0.00009} & \texttt{+0.00005} & \texttt{+0.00002}\tabularnewline
\hline 
$\mathcal{C}_{3}$-$S_{3}$/ lower & \texttt{-0.00009} & \texttt{-0.00005} & \texttt{-0.00003} & \texttt{-0.00002} & \texttt{-0.00002} & \texttt{-0.00003} & \texttt{-0.00006}\tabularnewline
\hline 
$\mathcal{C}_{3}$-$S_{3}$/ upper & \texttt{+0.00009} & \texttt{+0.00005} & \texttt{+0.00003} & \texttt{+0.00003} & \texttt{+0.00003} & \texttt{+0.00001} & \texttt{+0.00026}\tabularnewline
\hline 
$\mathcal{C}_{4}$-$S_{4}$/ lower & \texttt{-0.04279} & \texttt{-0.02138} & \texttt{-0.01069} & \texttt{-0.00428} & \texttt{-0.00215} & \texttt{-0.00109} & \texttt{-0.00044}\tabularnewline
\hline 
$\mathcal{C}_{4}$-$S_{4}$/ upper & \texttt{+0.04279} & \texttt{+0.02138} & \texttt{+0.01069} & \texttt{+0.00428} & \texttt{+0.00214} & \texttt{+0.00107} & \texttt{+0.00043}\tabularnewline
\hline 
\end{tabular}

\caption{\label{tab:gamma2}The difference of the total and dynamical phases
in the units of $\pi$ at the end of the paths $\left(t=T\right)$.
Different contours and initial wave functions are applied at $I=1\times10^{13}\frac{W}{cm^{2}}$
.}
\end{table}

\section{Conclusions }

By applying adiabatic time-dependent framework and Floquet representation
for the Hamiltonian we have calculated the geometric phase of the
light-induced conical intersection formed in the $\mathrm{D_{2}^{+}}$
molecule. It has been demonstrated that assuming certain conditions
for the initial wave functions the adiabatic time-dependent results
for the geometric phase are similar to those obtained from the time-independent
solutions \cite{LICI1,LICI2}. Obviously, obtained numerical results
are also in full agreement with the values of the Berry phase that
hold for the natural conical intersections. 

In the future, our aim is to compute the Berry phase for the exact
time-dependent light-matter Hamiltonian, too. However, this is not
an easy task because of the explicit time-dependence of the Hamiltonian.
The latter gives rise to additional difficulties and the adiabatic
transport round a close path is far from being trivial.

\section*{Acknowledgements}

The supercomputing service of NIIF has been used for this work. This
research was supported by the EU-funded Hungarian grant EFOP-3.6.2-16-2017-00005.
The authors thank Tamás Vértesi for many fruitful discussions.

\end{document}